\documentclass{ws-ijmpa}
\usepackage[super,compress]{cite}
\usepackage{graphicx}

\begin{document}

\markboth{V. N. Rudenko, Yu. M. Gavrilyuk, A. V. Gusev et al.}{Gravitational wave detector OGRAN as multi-messenger project of RAS-MSU}

%
\catchline{}{}{}{}{}
%

\title{Gravitational wave detector OGRAN as multi-messenger project of RAS-MSU}

\author{V. N. Rudenko$^{1,2}$, Yu. M. Gavrilyuk$^{2,1}$, A. V. Gusev$^{1}$, D. P. Krichevskiy$^{3,1}$, S. I. Oreshkin$^{1}$,\\ S. M. Popov$^{1}$, I. S. Yudin$^{4,1}$}

\address{$^{1}$Sternberg Astronomical Institute, Lomonosov State University, Universitetskii prospect 13, Moscow, 119234, Russia\\
$^{2}$Baksan Neutrino Observatory of INR RAS, Neutrino village, Elbrus district, 361609, Kabardino-Balkar Republic, Russia\\
$^{3}$Bauman Moscow State Technical University, Physics Department, 2-nd Baumanskaya street 5, building 1, Moscow, 105005, Russia\\
$^{4}$Moscow Institute of Physics and Technology, Institutskiy per. 9, Dolgoprudny,  141701, Moscow Region, Russia\\valentin.rudenko@gmail.com}

\maketitle

\begin{history}
\received{Day Month Year}
\revised{Day Month Year}
\end{history}

\begin{abstract}
Modernized version of the combined opto-acoustical gravitational wave detector OGRAN is presented. Located in the deep underground of the Baksan Neutrino Observatory this setup is aimed to work on the program of collapsing stars searching for in multi-channel manner with the neutrino telescope BUST. The both instruments have the sensitivity allowing a registration of collapses in our Galaxy as rare events with the estimated probability $0.03 $  year\textsuperscript{-1}. The OGRAN narrow band sensitivity at the kilohertz frequency is limited by its acoustical mode thermal noise achieving  $10^{-20}$ in term of  metric perturbations. A possible algorithm of the joint data analysis for the both instruments is developed and resulting formulae of the right detection probability are given. A future increasing of the OGRAN sensitivity associated with the moderate cooling (nitrogen temperature) of the acoustical mode is also discussed.

\end{abstract}

\ccode{PACS numbers: 04.08.Nn, 95.55.Vj}

\section{Introduction}	

 The first registration (September 2015)  of the gravitational wave (GW) signal from the merger of a binary formed by two black holes with masses of order $ 30{ M }_{ \odot  } $, from a distance of $ 400 $ Mpc was a landmark event, confirming the reality of hopes for gravitational wave astronomy as a new unique information channel in our  knowledge of the universe\cite{PhysRevLett.116.061102}.  The signal recorded by LIGO detectors in September 2015 amounted to $ h \simeq 10 ^{- 21} $ in terms of the dimensionless amplitude of the gravitational wave (or spatial metric variations in geometric language) at an average carrier frequency of $ 150 $ Hz.  In form, the signal fit into the well known portrait of so called “chirp signal” i.e. looks as a radio-pulse  with a changing carrier frequency. It has the three characteristic phases: inspiral, merging and relaxation (ring down). Until the end of 2016, a couple more similar signals were registered\cite{PhysRevLett.118.221101}. In 2017, two more significant events took place.  Firstly, the registration of signals from the merger of massive black holes performed already by three detectors.  The LIGO interferometer was joined by the European VIRGO interferometer, which increased the accuracy of the localization of the source of GW in the celestial sphere by a hundred times\cite{PhysRevLett.119.141101}.  Secondly, the same detectors recorded a signal from the merging of the components of a binary neutron star GW170817, which was accompanied almost simultaneously by the gamma ray burst GRB170817A\cite{PhysRevLett.119.161101} recorded by the Fermi\cite{2017GCN.21520....1V} and Integral\cite{Savchenko_2017} spacecrafts.  The 10 hours later, in zone of  localization of source of gravitational radiation a supernova belonging to the Galaxy NGC was discovered  by optical telescopes (including the MASTER  - global network of robotic telescopes\cite{Lipunov_2017}). This confirmed the hypothesis about the general nature of astrophysical sources of gravitational and gamma-ray bursts associated with relativistic catastrophes of super dense stars.  In this case, it was a merging of neutron components of the relativistic binary at the end of its evolution, which was also accompanied by a supernova explosion. Naturally, this achievement gave rise to a new surge in the activity of involved scientific collaborations and individual groups with a number of proposals for new long-term projects for high-sensitivity detectors, such as Einstein Telescope\cite{Punturo_2010}, Voyager and Cosmic Endower\cite{T1500290}.        
 
 LIGO and VIRGO instruments have achieved an extraordinary sensitivity to small  perturbation of their baseline (spectral density of  deformation noise $ {10}^{-21}{-}  {10}^{-22}$ Hz\textsuperscript{-1/2}) in the region from $10$ Hz up to $2$ kHz.  From the technical point of view these setups are extremely complex systems of automatic control of operational regime with a high level of seismic isolation, high vacuum in big volumes, modern laser physics technology etc\cite{RevModPhys.86.121} . It is a hard problem to keep a long time continuous operation without losing the working point.  In contrast the ability for long time (through years) continuous operation is the important positive feature of resonance bar detectors. Having such property, a resonance GW detector of moderate sensitivity ${10}^{-19}{-} {10}^{-20}$ Hz\textsuperscript{-1/2} could be used in the program of “search for rare events” - catastrophic processes with relativistic objects in our  Galaxy and close environment (100 kpc) accompanied by different types of radiation. In fact at present a similar programs are carried out also with underground neutrino telescopes. It consists in a search for Galactic collapsing stars on the neutrino channel. In this paper, we describe the present status of opto-acoustical gravitational antenna  OGRAN\cite{bagayev2014a-high7102717} constructed for a “multi-channel mode” monitoring relativistic events and located  the deep underground of Baksan Neutrino Observatory (BNO) close to the  Baksan Underground Scintillator Telescope (BUST). 

\section{GW Detector OGRAN Modernization}
\subsection{Technical status of initial setup}

\begin{figure}[b]
\centerline{\includegraphics[width=12.0cm]{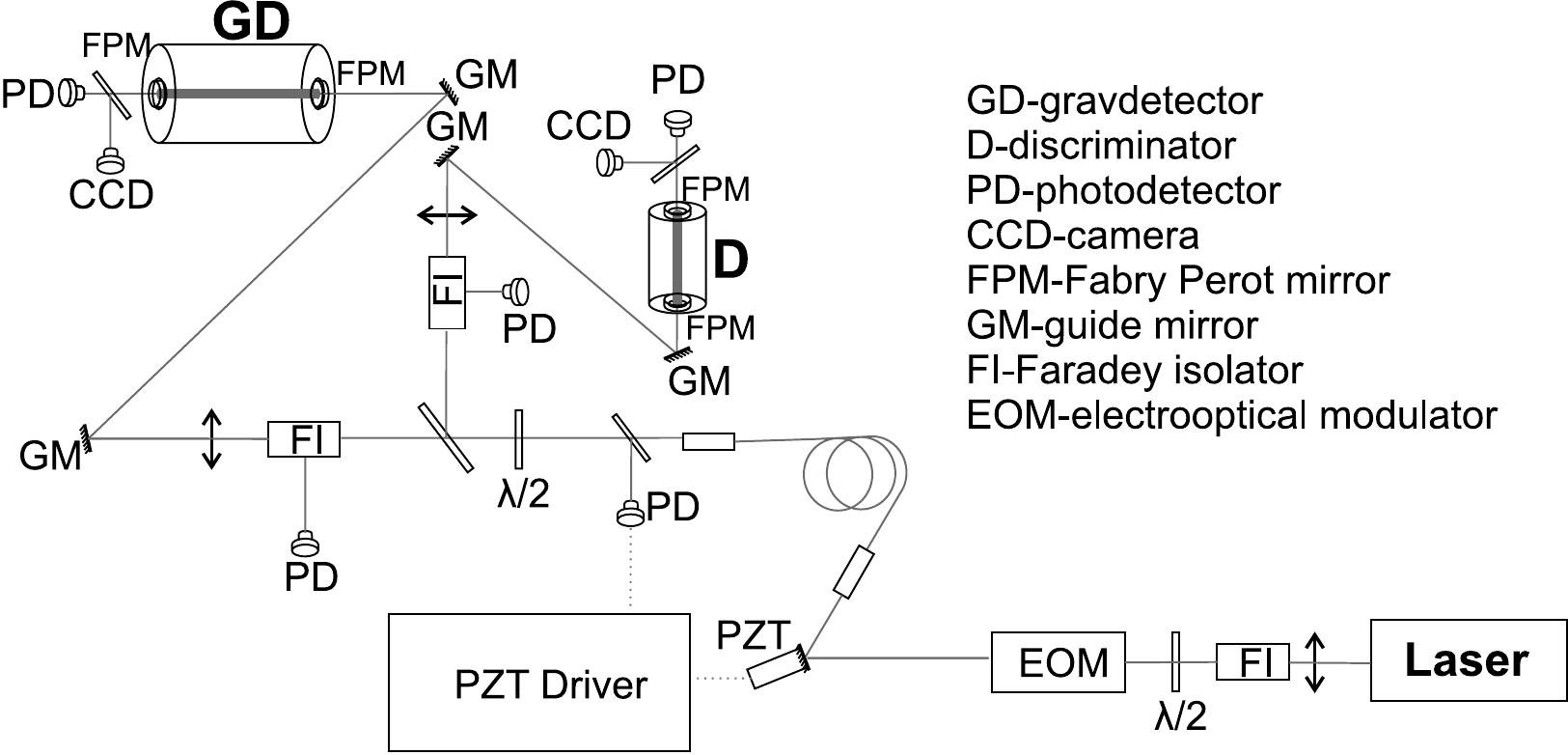}}
\caption{Principle scheme of the OGRAN antenna. \label{f1}}
\end{figure}

The idea of construction of a GW detector as an acoustical resonance bar coupled with the optical FP cavity composed by mirrors attached to the bar ends was considered in Refs. \citen{kulagin}, \citen{Bichak}. A new quality was emphasized of such a combination: (a) a more complex structure of signal response, containing separately acoustical and optical parts, and (b) a possibility to get sensitivity at the level of bar Brownian noise in a limited spectral frequency range due to the small back action of optical read out.  Implementation of this idea was reported as the OGRAN project. At present the full scale setup is constructed, tested and installed into underground facilities of BNO INR RAS.  The principal opto-electronic scheme of the setup is given in Fig. \ref{f1}. Generally, it belongs to the design type called a comparator of optical standards. It is composed by two feedback loops. The first one couples the Fabry-Perot cavity (FP) on the large bar (OGRAN acoustical detector) with the laser of optical pump which frequency automatically tuned in resonance with bar FP cavity. Any change of the FP optical length is converted into the pump beam frequency variation. The second loop is the measuring one. It consists of the additional FP cavity (the frequency discriminator) illuminated by the same pump laser but tuned in resonance by an independent piezo-ceramic driver attached to one of discriminator’s mirror. Its output signal is proportional to the frequency difference between of laser pump and of discriminator's cavity. Any perturbation of the detector bar cavity (above of feed back cut off frequency $\sim 100$ Hz) is reflected in the output discriminator signal. It worth to emphasize once more the OGRAN physical specificity: the gravitational wave interacts not only with acoustical degree of freedom (the resonance bar) but with the EM field in the cavity as well. It results in a complex structure of signal response with optical and acoustical parts. A payment for this originality is the technical problem of constructing a large scale high finesse FP cavity rigidly coupled with the acoustical resonator without loosing its high mechanical resonance quality factor Q.

The initial version of the OGRAN antenna was described in details in the Ref. \citen{bagayev2014a-high7102717}. It had the following physical and technical characteristics: acoustical detector effective mass $M = 10^3$ kg, length $L = 2$ m, acoustical resonance frequency ${ \omega  }_{ 0 }/2\pi=1.3$ kHz under the $Q={ 10 }^{ 5 }$. Parameters of optical part were the following: laser wave length $1064$ nm, laser total power  $W\approx 2$ W, effective power in the FP cavities $ 0.05 $ W, finesse: $F=3000$ of detector cavity, $F=1500$ of discriminator cavity, contrast of the interference $C\approx 0.2$. The experimental sensitivity curve in the form of noise spectral density was presented in Ref. \citen{bagayev2014a-high7102717}. It resulted in the minimal detectable strain ${ (\Delta L/L) }_{ f }\approx { 10 }^{ -18 }$ Hz\textsuperscript{-1/2} in the bandwidth  $\Delta f\sim 30$ Hz,  and  ${ (\Delta L/L) }_{ f }\approx { 10 }^{ -19 }$ Hz\textsuperscript{-1/2} in the bandwidth  $\Delta f\sim 4$ Hz around of central receiver frequency $1.3$ kHz.

\subsection{Technical improvements of the installation}
\subsubsection{Seismic and acoustic isolation}
Despite the underground location of the OGRAN antenna, the residual mechanical vibrations of the environment negatively affected the stability of its functioning. The measuring part of the setup was most susceptible to their influence. First in that number is the frequency discriminator having the form of cylindrical FP cavity (made from the ULE optical glass  - sitall) located in a vacuum chamber on a massive optical table. Due to the fact that the discriminator is installed in a vacuum chamber in a vertical position, there are increased requirements for its seismic isolation. For these purposes, an anti-seismic suspension was developed, which includes a multi-link anti-seismic filter and acoustic sound insulation of a vacuum chamber. This antiseismic filter was composed by a system of several stainless steel rings separated by elastic tablets of isolating material (viton). The size of the tablets was selected from the calculation of the total load of the upper rings and the body of the discriminator. The lower tablets are larger in size. The conical system of rings was used as amore resistant to lateral vibrations of the inclined mode. In the upper link of the filter, a spring suspension with damping by soft mesh sponge made of stainless steel is applied to lower its eigen frequency. By selecting the size and rigidity of the dampers, as well as reducing the size of the viton tablets between the filter rings, we were able to reduce the eigen frequency of the entire filter to $3$ Hz with a quality factor of about $ 0.3 $. The transient response of the filter looks like a step with a slight rise at the $ 3 $ Hz resonance and subsequent decline. The steepness of the decline is determined by the eigen frequencies of the steps and their number.
To reduce the external acoustic impact, the discriminator's vacuum chamber was externally coated with a two-layer shell of sound-absorbing material such as Comfort Mat $ 10 $ mm thick.

\subsubsection{Optical and electronic parts}
\begin{figure}[b]
\centerline{\includegraphics[width=10.0cm]{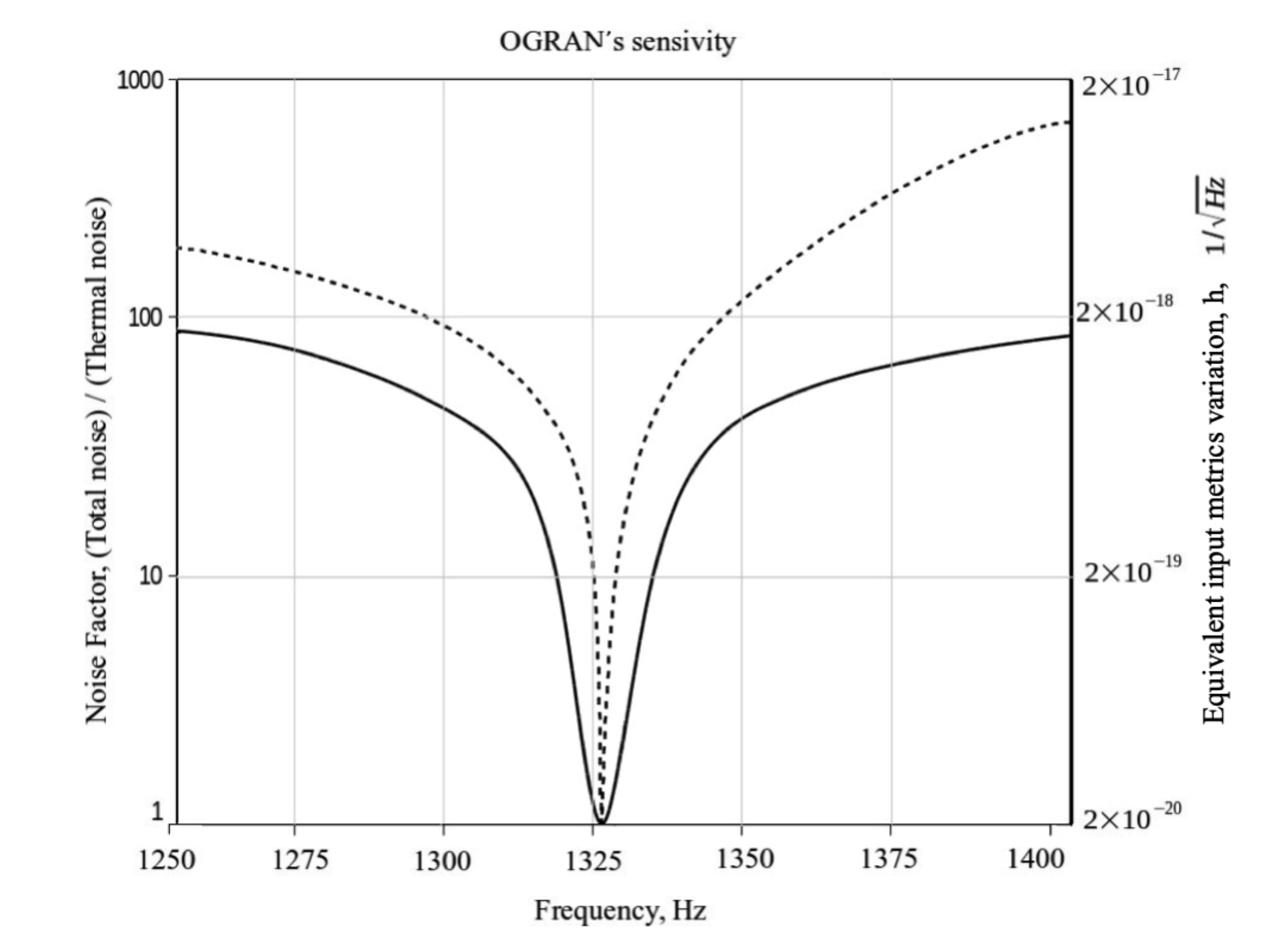}}
\caption{Noise spectral density of the OGRAN antenna
(doted line - the experimental curve of old version of the setup  $ F \sim  2000 $;
normal line - the calculated curve of modernized setup   $ F \sim  30000 $).\label{f2}}
\end{figure}

\begin{romanlist}[(ii)]
\item The optical parts of the OGRAN setup were drastically updated. The mirrors of both FP cavities were replaced by the high quality mirrors (manufactured  in the Laboratoire des Matériaux Avancés (LMA), Lion France). The new mirrors have the same geometry, but very low dissipative losses – $ 1-2 $ ppm of the absorptions and  $ 3-6 $ ppm of the scattering.  After mirror replacement the finesse of the FP of the OGRAN detector became $ \sim 30 000 $, reference cavity $ \sim 78 000 $, that lead to the increasing of amplitude transfer function coefficient more than one order of value. The contrast of interference was also increased up to $ 60 \% $.
\item The next step was done in the direction of excess noises depressing. One of the principle effect of this type is an undesirable (parasitic) amplitude modulation of the optical pump. Operating regime (operating point) of the OGRAN antenna is kept by the system feedback which works on radio frequency of $ 10.5 $ MHz. The appearance of this harmonic in the opto - electronic tract of antenna is provided by modulating of the optical pump radiation with the subsequent  selective photo detection. The principle of feed back circuits is based on a specific PDH technique\cite{Drever1983} which uses phase modulation of the light beam. With pure phase modulation in the PDH technique, the excess noise does not occur. However, the available phase electro-optical modulators due to their not perfect and inaccurate settings bring also the residual (spurious) amplitude modulation of the pump radiation. Detection of such radiation on the photodetectors generates excess noise in electronic circuits. The effect of residual amplitude modulation (RAM) in electro-optical phase modulators is not fully understood. Measurements have shown that it depends on the transverse inhomogeneity of the laser beam, as well as on the properties of the propagation medium\cite{Zhang:14}.  We used the empirical method of actively suppressing RAM at the output of the phase modulator. A ray passing through a modulator enters an optical fiber (that preserves the polarization of light) through a mirror (on a piezo driver).  At the fiber output, a small fraction of the light is cut off to the photo detector to generate an electronic error signal.  The latter is produced only by light components of amplitude modulation. The error signal goes to the driver of the mirror, which holds the position of the wavefront of the beam at the entrance of the optical fiber corresponded to the minimum (zero) error signal (lack of RAM).  Experimental results showed that after applying such an active feed back system, RAM noise is suppressed by more than $ 30 $ dB.

\item The magnitude of the output signal OGRAN is proportional to the optical pump power. However, the available photodetectors have limited power dissipation.  Also, when operating at high frequencies, the parallel shunt capacitance of the photodiode is important (for a modulation frequency of $ 10.7 $ MHz, the effect is significant).  There is a contradiction between the large working area that reduces the power density, and the small shunt capacitance of the diode.  By itself, working at a high power level leads to heating of the sensitive layer of photodiodes, which also increases their capacitance. In the improved OGRAN installation, silicon photodiodes in the old version are replaced by photodiodes from InGaAsP, which have the better quantum efficiency - $ 0.7 $ versus $ 0.4 $ at a wavelength of $ 1064 $ nm. In practice, commercially available photodiodes can operate efficiently at a power of $ 30-50 $ mW at frequencies of tens MHz.  At higher power, the signal at $10.7 $ MHz shunts with increasing power and stops growing.  It is possible to overcome this limitation when switching to multi-channel detection for high powers  $ \ge 1 $ W, where each photodiode serves its own fraction of the light. With an improved version of OGRAN, preliminary experiments were carried out with $4$ and $16$ channel detector arrays using light dividers.  In practice, it is difficult to obtain equal optical power in each beam and tune all channels to maximum efficiency.  It is planned to test fiber-optic photo detectors for this task. Only the most important enhancements implemented in the new version of OGRAN are listed above. The most significant is the modernization of key elements - Fabry-Perot interferometers in both arms of the antenna. The replacement of ordinary mirrors with super-reflecting high-tech mirrors with vanishingly small losses.  In this case, the conversion coefficient of the signal perturbation from the gravitational-deformation form to the optoelectronic one is increased more than an order of magnitude. As a result, the band of effective reception of GW signals at the OGRAN antenna expands in proportion to the square root of the FP finesse ratio.  This is illustrated in Fig. \ref{f2}, which shows the calculated spectral density of the deformation noise for a modernized antenna in comparison with its initial version (experiment described in in Ref. \citen{bagayev2014a-high7102717}).  At the level of $ { 10 }^{ -18} $ Hz\textsuperscript{-1/2}, the reception band expanded from $ \sim 7 $ Hz to $ \sim 30 $ Hz.
\end{romanlist}

\section{OGRAN Potential and Factual Sensitivity }

Let’s come back to the calculation of sensitivity of the OGRAN antenna , considering the both natural type of noises: the brown noise of its acoustical resonator and photon short noise of the optical one. For simplicity we will neglect of the complex structure of the antenna reaction, containing of acoustical and optical parts in its response to GW . To estimate sensitivity it is sufficient to follow the simple scheme in which the GW signal excites of the fundamental mode of bar acoustical resonance but the optical FP cavity serves as a read out system.

At the output of such read out one has some bar length displacement $x(t)$ produced by equivalent GW force ${{f}_{g}}(t)$ together with the Langevin force of thermal oscillation ${{f}_{T}}(t)$. The oscillation $x(t)$ appear also at the additive read out (optical) noise back ground ${{x}_{a}}(t)$. Thus the total output displacement ${{x}_{tot}}(t)$ in the complex amplitude formalism is presented as 
$${{x}_{tot}}(t)=K(p)\left[ {{f}_{g}}(t)+{{f}_{T}}(t)+{{K}^{-1}}(p){{x}_{a}}(t) \right],$$
where the oscillation mode transfer function $K\left( p \right)=[M{{({{p}^{2}}+2\delta p+{{w}_{0}}^{2})]}^{-1}}$.

The theory of optimal filtering suggests to perform the signal extraction coming back to the antenna input i.e. applying the inverse filter. Then the spectral density of the noise background after the inverse filter reads as
$$N(\omega )={{N}_{T}}(\omega )+K(j\omega ){ }{N}_{a}(\omega ).$$
Here ${{N}_{T}}(\omega )$, and ${N}_{a}(\omega )$ are the spectral densities of thermal fluctuation and additive read out noise. Below they are considered as constant ${{N}_{T}}$, ${{N}_{a}}$.

Considering the processes in a narrowband region around of the fundamental acoustical mode resonance frequency ${\omega }_{0}$ and supposing the smallest dissipation index $\delta << {{\omega }_{0}}$ one comes to the approximation
$${{N}_{0}}(\omega ) = N({\omega }_{0}+\omega )\approx {{N}_{T}}+ (2m{{\omega }_{0}})2{{N}_{a}}{{\omega }^{2}}+ o({{\delta }^{2}}), |\omega |<<\omega_{0} .$$ 
For estimation of the potential and real OGRAN sensitivity it is convenient to present this noise background introducing a new parameter - the “noise factor” $F$:
$${{N}_{0}}(\omega ) = {{N}_{T}}[ 1 + {{N}_{a}}{{(2M{{\omega }_{0}})}^{2}}{{\omega }^{2}}/ {{N}_{T}}] =~{{N}_{T}}F(\omega ).$$
It is convenient to rewrite the noise factor in a more practical form:
$$F(\omega )=1+{{(\omega /\Delta \omega )}^{2}}, {{(\Delta \omega )}^{2}}= {{N}_{T}}/({{N}_{a}}{{(2M{{\omega }_{0}})}^{2}}).$$
Here $\Delta \omega $  is the effective bandwidth of potential sensitivity: in the region around the antenna resonance frequency ${{\omega }_{0}}\pm \Delta \omega $ where the sensitivity is limited only by the thermal fluctuation of the acoustical degree of freedom.

Let’s apply these general formulae directly to the OGRAN setup, using the parameters of the old version presented in Ref. \citen{bagayev2014a-high7102717}. The minimum metric perturbation registered by the opto-acoustical detector using the optimal filtering procedure is read as
$${{h}_{min}}= (\Delta L/L)\approx \left( 4/L \right){{[(kT/M{{\omega }_{0}}) (1/Q/{{\omega }_{0}}\tau )]}^{1/2}}\sqrt{F}.$$
The potential sensitivity corresponds to the noise factor \(F=1\). Substitution of the OGRAN parameters: $M={{10}^{3}}$ kg, $L=2$ m, ${{\omega }_{0}}=8*{10}^{3}$ rad/s, $Q=1.2*{{10}^{5}}$ and the duration of GW burst $\tau \approx (1/\Delta f)=2*{{10}^{-3}} $ s ($2$ periods of resonance frequency) results in the sensitivity estimate ${{h}_{\min }}=1.6*{{10}^{-18}}$. The correspondent receiver band width is $500  $ Hz. In the more short bandwidth the sensitivity forecast is at the level ${{h}_{\min }}=2*{{10}^{-19}}$. At last in the very narrow bandwidth $\sim 1 $ Hz it can reach ${{h}_{\min }}=5*{{10}^{-20}}$. 

It is known that the relativistic catastrophic events type of SN explosion can provides $h={{10}^{-18}}$ from the distance $10 $ kpc if the part of energy contributed in GW will be $1 \% $  of the solar mass. However the number of papers with calculation of the collapse efficiency forecast GW pulse signal in the best case at the level $h\le {{10}^{-20}}$. It means that the reachable sources would be located at more close distances $\sim 1 $ kpc, i.e. the GW collapse detection with OGRAN belongs to the “search of rare events” programs.

As for condition of the “small noise factor” $F\le 1$ for the opto-acoustical antenna a one can say that it was satisfied already in the first version of the OGRAN detector with the optical pump power $\sim 0.1 $ W and FP cavity finesse $F=2000$. Modernized OGRAN version uses the pump power $2 $ W and has the finesse in the detector FP resonator $F=30000$, i.e. the requirement $F<1$ is fulfilled with a large margin. 

\section{Neutrino and Gravitational Radiation of SN}
During of supernova flare, an energy of $ E\approx 3\times { 10 }^{ 53 } $ erg is released.  Neutrinos carry out $\sim 0.99E $; the characteristic energy of runaway neutrinos is $\sim 10 $ MeV.  In accordance with the standard theory of supernova explosion\cite{1985ApJ...295...14B}, it is the neutrino collapse mechanism that causes the shock wave of the rebound and, therefore, the expansion of the outer layers.  There are two main processes in which neutrinos are produced\cite{1987ApJ...318..288M}: 
\begin{romanlist}[(ii)]
\item capture of electrons by protons of the iron nucleus:  ${ p }^{ + }+{ e }^{ - }\rightarrow n+{ \nu  }_{ e } $; neutrino radiation occurs at initial stage with a huge luminosity $L\sim { 10 }^{ 53 } $ erg/s, during of a very short time:     $\tau \lesssim 0.1$ s; in this process   $ 5 \% $ of all neutrinos emitted during the collapse is released; 
\item annihilation reaction: ${ e }^{ + }+{ e }^{ - }\rightarrow { \widetilde { \nu  }  }_{ i }+{ \nu  }_{ i } $, where $i=\mu ,\tau ,e $ ; here the neutrino emission occurs during of a time $\tau \sim 10$ s with luminosity $L\sim { 10 }^{ 52 } $ erg/s.  Just in this process the electronic antineutrinos are released that can be detected by BUST\cite{Novoseltsev2017}. In the program of searching for a neutrino burst from SN, the BUST sensitivity radius is approximately $20 $ kps.
\end{romanlist}

GW radiation from SN\cite{PhysRevD.78.064056} occurs during a nuclear rebound due to the non-spherical collapse (there is a third derivative of the quadrupole moment).  The expected value of the dimensionless GW amplitude, which came from a distance of $\sim 10$ kpc, is $h\lesssim { 10 }^{ -20 } $.  The characteristic frequency of the wave packet is   $f\sim 1000$ Hz.  Depending on the model parameters (star mass, degree of non-spherical form factor of the collapse), the GW energy is in the range from ${10}^{44}$ to ${10}^{49}$ erg.  As it was shown in the Section 2 this forecast is a bit below of OGRAN sensitivity in its present version. 

BUST\cite{Novoseltsev2017} is located at BNO RAS an effective depth of $850 $ m.  The setup consists of $3184$ scintillation counters, the total mass of the scintillator is $330 $ t. The scintillation counter is an aluminum container $0.7\times 0.7\times 0.3$ m\textsuperscript{3} in size, filled with a scintillator based on white spirit (${ C }_{ n }{ H }_{ 2n+2 }, n \approx 9$). 

Most of the events that BUST records from a supernova explosion (SN) are the result of reverse beta decay:  ${ \bar { \nu  }  }_{ e }+p\rightarrow n+{ e }^{ + }. $   At the average antineutrino energy of $12-15 $ MeV, the range of length of the positron born in such reaction will be enclosed in the volume of one counter.  Thus, the search for a neutrino burst consists in registering a cluster (group) of single events during the time interval $ \tau = 20 $ s - a typical duration of a neutrino burst from SN.  For SN at a distance of $10 $ kpc, the total energy emitted into the neutrino is $3\times { 10 }^{ 53 }$ erg and the target mass is $130 $ t (three lower BUST planes), the estimate of the expected number of events from a single collapse is ${ N }_{ \nu  }\approx 35 $. Of course, there is a random background of such events created by the radioactivity of the environment, muons of cosmic rays, false alarms of counters, etc. The background is such that it creates a cluster of $ k=8 $ single events at a rate of $0.138$ year\textsuperscript{-1}. In $ 10 $ years, no more than $ 2 $ events can be expected.  The cluster formation rate from $ k = 9 $ background events is already falling significantly and is equal to $7\times { 10 }^{ -3 }$ year\textsuperscript{-1}.  Thus, clusters with $k>9 $ cannot be created by the background and are candidates for SN flare registration.  The measurement method adopted in the experiment consists in recording events on a moving $20$ s time interval that moves from one single event to the next, so that at least one event is always present in the cluster.

\begin{figure}[b]
\centerline{\includegraphics[width=10.0cm]{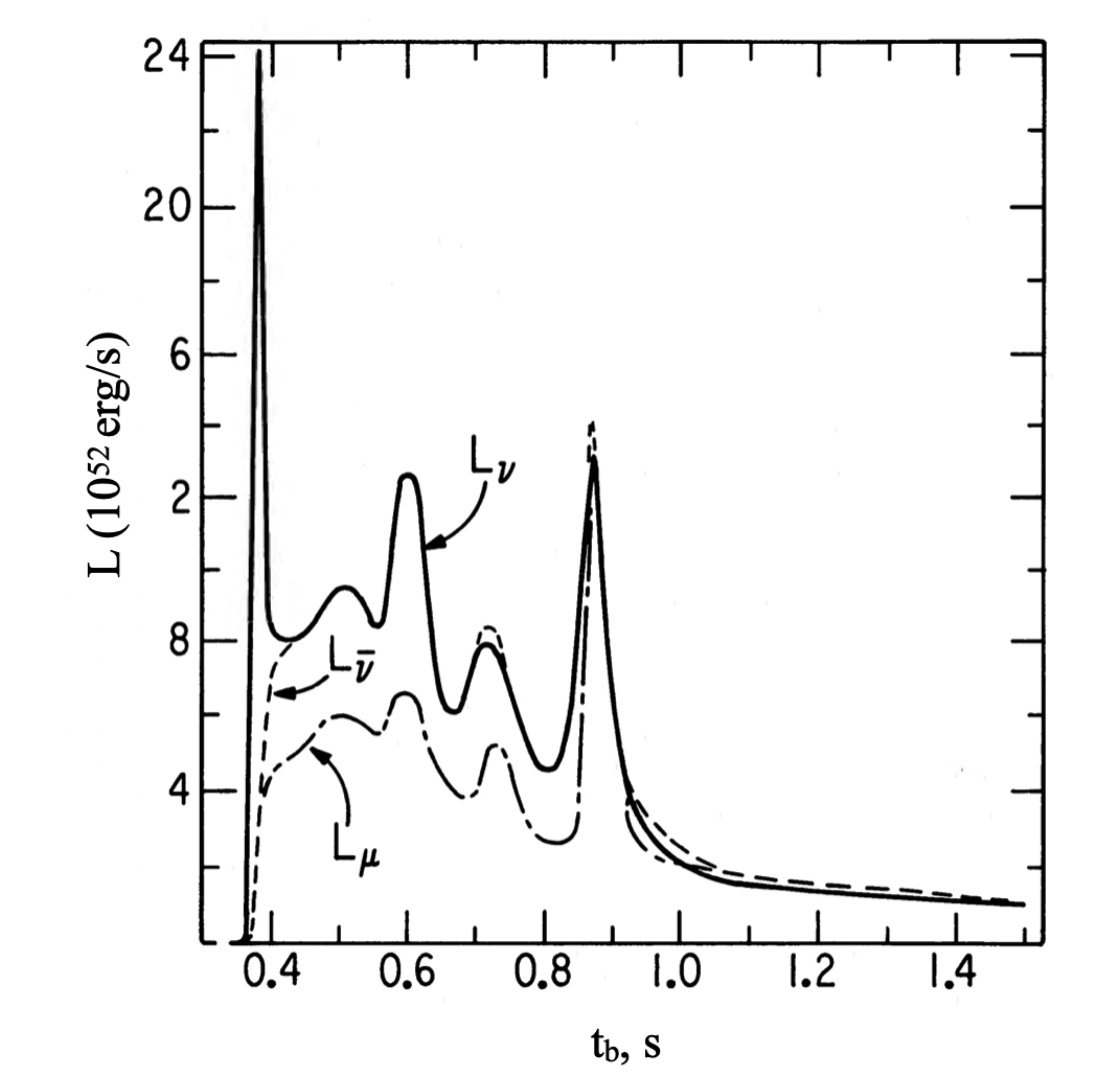}}
\caption{Neutrino luminosity of collapsing star $M=20 {M}_{\odot }$, ${t}_{b}$ $-$ time after bounce (R.Mayle,  J.R.Wilson  and  D.N.Schramm, \textit{Astrophys.J.} \textbf{318},  288  (1987)).
\label{f3}}
\end{figure}

\begin{figure}[b]
\centerline{\includegraphics[width=10.0cm]{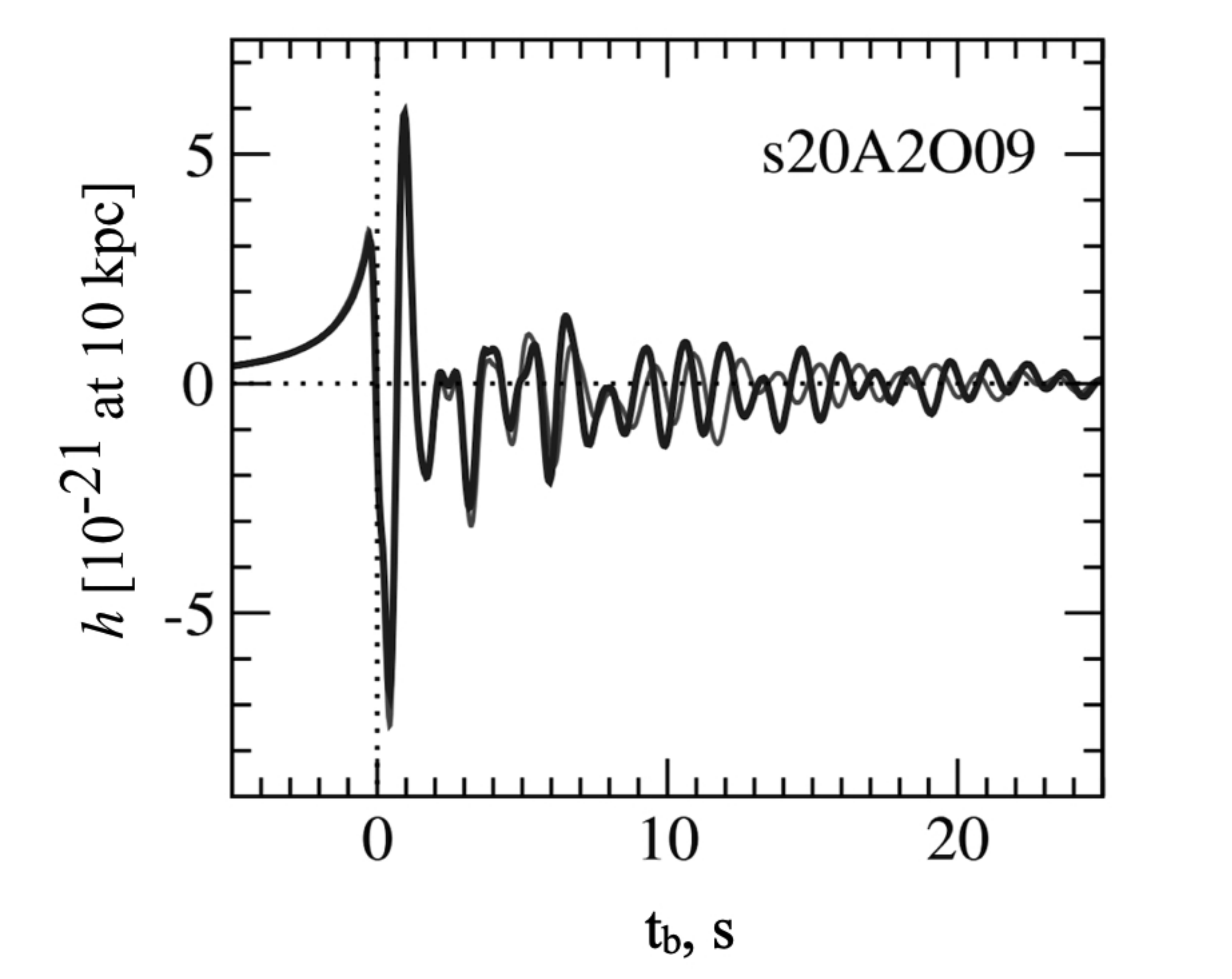}}
\caption{Structure of GW burst from collapse of rotation star $M=20  {M}_{\odot  }$, ${t}_{b}$ - time after bounce (H.Dimmelmeier,  C.D.Ott,  A.Marek  and  H.T.Janka, \textit{Phys.  Rev.  D} \textbf{78},  064056  (2008)).
\label{f4}}
\end{figure}

Numerical computations of the both neutrino and GW bursts are presented on the Fig. \ref{f3} and Fig. \ref{f4}. One can note in neutrino luminosity curve (Fig. 3) the large initial peak and following after more weak peaks reflecting radiation of collapse bounces. According to the papers (Refs. \citen{1987ApJ...318..288M}, \citen{Janka2015}) all such processes are developing inside of very short time on the order on $\sim 1$ s. However there are other scenarios of multi - stage collapse\cite{Imshennik2004, Bisnovatyi_Kogan_2017} forecasting of multi-radiation flux during of a total collapse duration $\sim 20$ s. In our analysis below we take it into account.  The numerical modeling of GW radiation from non-spherical collapse\cite{Imshennik2004,Bisnovatyi_Kogan_2017} gives of the enough complex pulse form (Fig. 4). But in this paper for simplicity we use an approximation of more intensive part of the burst as a very short radio pulse with a smooth Gaussian envelope and a resonant carrier ($1-3$ periods).

\section{Strategy of Two Channel Search for Collapsars}
The search for collapsars at the BUST neutrino telescope has been conducted for over $25$ years. The result is an experimental estimate of the upper boundary of the frequency of occurrence of such events in the Galaxy $\sim 0.03$ year\textsuperscript{-1}. By connecting the OGRAN gravitational antenna to the observations, it is potentially possible to increase the search efficiency. In this regard, it is important to develop a procedure for joint data processing of both tools.  Below there  is a variant of such processing.  Its principle is to improve the detection characteristics of the gravitational antenna due to additional information from the neutrino telescope about the current background of neutrino events. In turn, the control record of the gravitational detector can further reduce the number of candidates (suspicious bursts) for signal neutrinos from collapses.  Specifically, the analysis of the neutrino background allows us to narrow the time interval for the search for GW - disturbances on a gravitational antenna. Let's explain it more in detail.

\subsection{Description of GW detector data processing}
Let's consider random process on the OGRAN output:
\begin{equation}
x(t)=\lambda s(t)+n(t),\
\label{diseqn1}
\end{equation}
where $ \lambda = (0;1)$ - detection parameter, $ s(t) $ - signal response, $ n(t) $ - additive Gaussian noise with known spectral density  $N(\omega)$. 
For a short GW-burst the equivalent force at the detector input formally can be 
presented as: $ { F }_{ s }(t)=A\delta (t-\tau )$; where $A$ - unknown amplitude, $\tau $ - unknown moment of arrival time with uncertainty in a  prior interval $ ({ \tau  }_{ min };{ \tau  }_{ max })$. The additive mixture $x(t)$ is processed according to the optimal filtering scheme
\begin{equation}
x(t)\rightarrow MF\rightarrow AD\rightarrow R(t)\rightarrow \max { R(t_{ 0 }+t) } \ge { c }_{ \alpha  }\ for\ t\in ({ \tau  }_{ min };{ \tau  }_{ max }),\ 
\label{diseqn2}
\end{equation}
where the blocks of processing marked as:  $MF$ - matched filter; $AD$ - amplitude 
detector getting an envelope of a narrow-band realization at the output of $MF$, 
${t}_{0}$ - time delay introduced by $MF$, $ { c }_{ \alpha  }$ - threshold lever of Neyman-Pearson's receiver\cite{Levin} defined by false alarm error $ \alpha $ (an exceeding of threshold means a presence of the gravitational signal or $\lambda =1$).
Let  $ \sigma ^{ 2 }$ is the variance of Gaussian noise at the $MF$-output,  $ \Delta f$ - the bandwidth;  then in first approximation
\begin{equation}
{ c }_{ \alpha  }\approx \sigma \sqrt { 2\ln { \frac { B }{ \alpha  }  } } ,\ 
\label{diseqn3}
\end{equation}
where the new dimensionless parameter $B$ is introduced so that $B\approx ({ \tau  }_{ min }-{ \tau  }_{ max })\Delta f$. 
The physical information of a close time position of neutrino and GW signals (neutrino-gravitational correlation) allows us to reduce the prior interval of searching $({ \tau  }_{ min },{ \tau  }_{ max })$, choosing  $\left| \tau -t_{ c } \right| <1\ s$; where $t_{ c }$ – the start moment of collapse. It decreases of the threshold ${ c }_{ \alpha  }$  and corresponding value of  $R(t_{ 0 }+t)$.

\subsection{Principle of joint data processing}

In our paradigm of neutrino gravitational correlations, the probability of
missing of gravitational signal on GW detector output  $ { \beta  }^{ * } $ is composed by two terms
\begin{equation}
{ \beta  }^{ * }=\beta +{ p }_{ e }(1-\beta ),\
\label{diseqn4}
\end{equation}
where the first is $\beta$ - probability of missing of GW signal with unknown arrival time $ \tau $  at the selected observational interval, and the second contains a $ { p }_{ e } $ - probability of error in choosing of this interval, which is defined as
\begin{equation}
{ p }_{ e }=\alpha +{ p }_{ m }(1-\alpha ),\
\label{diseqn5}
\end{equation}
where $ \alpha $ - false alarm rate in the neutrino channel, $ { p }_{ m } $ - probability of error in choosing of the maximum neutrino signal which is calculated below.
By taking into account expected duration of gravitational collapse $ { T }_{ 0 }\approx 20$ s we use a neutrino rolling counter which counts neutrino events at following
$ { T }_{ 0 }=20$ s intervals  $ \Delta n(t)=n(t)-n(t-{ T }_{ 0 }) $:
if the number of neutrino events at some interval $ (t-{ T }_{ 0 },t) $ exceeds the
threshold  level $ \Delta { n }_{ 0 }=\Delta n(t_{ 0 })>\Delta { n }_{ \alpha  } $; then one considers of  possible appearing of GW signal on this interval.
Supposing of Poissone low for the neutrino particles one can use of general formula of 
probability to get no more the $m$-counts\cite{Levin} for the pulse flux with average velocity of counting  $ \left< m \right> $:
\begin{equation}
P(m;\left< m \right> )=1-\frac { \Gamma (m+1,\left< m \right> ) }{ \Gamma (m+1) } ,\
\label{diseqn6}
\end{equation}
where $ \Gamma (a,z) $ - incomplete Gamma-function, $ \Gamma (a) $ – Gamma function.
The quite (standard) condition of neutrino detector corresponds to the following relation
\begin{equation}
1-\alpha =P(\Delta { n }_{ \alpha  }+1;\left< \Delta { n }_{ 0 } \right> ),\
\label{diseqn7}
\end{equation}
where $ \left< \Delta { n }_{ 0 } \right> $ - average number of noise neutrino events on observation interval. Now one can consider fine structure of neutrino events on selected individual interval. Let's divide the random process $\Delta n(t_{ 0 }-{ T }_{ 0 }\le t\le t_{ 0 })$ into sub-intervals with length, say: $ \Delta t\approx 2 $ s ($10$ sub-intervals). On each sub-interval there are ${m}_{i}$ events. From the set of sub-intervals one selects of those which satisfy of the local maximum condition:
$$ \Delta { m }_{ i }\ge \Delta { m }_{ \alpha  };\ \Delta { m }_{ i\pm 1 }<\Delta { m }_{ i },$$
where $ \Delta { m }_{ \alpha  } $ - threshold level for sub-intervals. This procedure reflects the presents of local maximums on the theoretical picture of neutrino luminosity (Fig. 3) corresponding to core collapse bounces. In that case we can define $\alpha$ as
$$ 1-\alpha =P(\Delta { m }_{ \alpha  };\left< \Delta { m }_{ 0 } \right> ), $$
where  $ \left< \Delta { m }_{ 0 } \right> $ - average number of noise neutrino events on sub-intervals. At last the probability $ { p }_{ m } $  required above is calculated according to the formula
\begin{equation}
{ p }_{ m }=[1-P(\Delta { m }_{ i }+1;\left< \Delta { m }_{ i } \right> )]P(\Delta { m }_{ i+1 };\left< \Delta { m }_{ 0 } \right> )P(\Delta { m }_{ i-1 };\left< \Delta { m }_{ 0 } \right> ).\
\label{diseqn8}
\end{equation}
Formulas (3), (4), (7), (8) present the statistical estimate of registration of  a collapsing  object through the joint processing of two channel data provided by neutrino and GW detectors. 

A more detailed comment is as follows. Formula (4) gives the probability of missing of the gravitational signal at the threshold level ${ c }_{ \alpha  }$ (3) with optimal processing (2) of the  gravitational antenna output obtained in the time interval $({ \tau  }_{ min };{ \tau  }_{ max })$ prompted by the BUST neutrino detector data (in vicinity of the suspicious neutrino reference).
Above the  probability of false alarm (chance of occurrence) ${ c }_{ \alpha  }$  due to the  thermal noise for OGRAN and Poisson background for BUST was introduced. It is reasonable to accept the same value in both cases without serious damage to the generality of analysis.  In fact, both statistics affect the solution to the detection of a multi-channel event: the thermal one - to set the registration threshold (3), Poisson one - to optimize the choice of the search interval of gravitational bursts. 
To test this technique, a computer simulation is planned with the injection of artificial (calibration) signals into both detectors.

\section{Discussion of Results and Development Prospects}
In the first part of this article the new qualities of the GW antenna OGRAN after its modernization were described.  The main thing is the expansion of the reception band due to the increased finesse of FP resonators.  But the ultimate sensitivity still cannot exceed the level ${ 10 }^{ -20 }$ in metric perturbations (limited by the thermal Brownian noise of the acoustic detector). The radical way to overcome this barrier is associated with deep cooling (up to $\sim 10$ K) of the acoustic resonator.  This expensive enterprise requires a significant increase in funding from the experience of cryogenic detectors Explorer and Nautilus\cite{PhysRevD.76.102001}.

However, in Ref. \citen{Kulagin2016} an intermediate order of magnitude less costly option was proposed, associated with cooling the acoustic detector to a nitrogen temperature of $\sim 80$ K. The vacuum chamber of the OGRAN detector can relatively easily be transformed into a nitrogen cryostat as a result of adding an internal nitrogen bath and enveloping the detector with a thermally insulating  screen shell.  The corresponding design was developed and created for the cryo-OGRAN pilot model\cite{Kvashnin17}.

The results of test experiments\cite{Kvashnin17} on this setup do not refute the possibility of increasing the sensitivity of such a “nitrogen version” of the antenna to a level of $\sim 3\times { 10 }^{ -21 }$ Hz\textsuperscript{-1/2}.  This already corresponds to the comparable radii of the location zone of detected collapses (supernova explosions) along both observational channels of neutrino and gravitational.

It is the place to recall a well known case of detecting the neutrino and gravitational signals from SN1987A\cite{Aglietta_1987, ALEXEYEV1988209}. This was the only example of a two channel correlation between the signals detected by a neutrino telescope and by bar gravitational wave detectors with piezo sensors operating at room temperature. Later on, only neutrino events were recognized as “the first detection” of a neutrino flux from a collapsing star\cite{PhysRevD.38.448}. However, in the process of analysis, some algorithm of a multi-channel detection was proposed and applied. That algorithm was based on the empirical estimates of the coincidence between the signals of different nature with arbitrary relative time shifts of both data sets\cite{Aglietta_1987}. The fact of the neutrino - gravitational correlation in the case of SN1987A was not confirmed\cite{PhysRevD.51.2644, Rudenko2000}, because the sensitivity of the Weber’s type resonance bar detector operating at room temperature was insufficient for detecting of signals of astrophysical origin. The opto-acoustical antenna OGRAN even without cooling has the sensitivity $2-4$ orders of magnitude greater than the Weber’s bar detectors. In the version cooled to nitrogen temperature, its advantage will reach five orders of magnitude in sensitivity.

As for the joint data processing algorithm for both channels, the algorithm described in Section $5$ should be more efficient than a simple “coincidence algorithm"\cite{Aglietta_1987,ALEXEYEV1988209} because it takes into account the fine structure of neutrino events in areas of records where neutrino anomalies are present.

In conclusion, we note the importance of the very problem of detecting neutrino-gravitational signals from collapse.  Here there is a much richer Physics than that contained in gravitational bursts (chirps) emitted during the merger of relativistic binaries.  In fact, the chirp structure in the inspiral stage is predicted quite well already in the framework of Newtonian theory and provides information on the parameters of the binary (mass, half-axis, frequency). The subtle relativistic details of the binary are not yet resolvable, as are nuclear processes in the merging stage. On the contrary, the temporal structure of neutrino and gravitational bursts from a collapsing star is just an indicator of the nuclear processes taking place in it\cite{1987ApJ...318..288M,Janka2015,Bisnovatyi_Kogan_2017}.  In particular, “bounces” in the monotonic compression course indicate a change in the equation of state of nuclear matter with increasing density, temperature, etc.  This argument is the main motive of the BNO INR RAS program for the two-channel search for collapsars in the Galaxy.

\section*{Acknowledgments}

The authors express their gratitude to the organizers of the Friedman Conference in June 2019: V. M. Mostepanenko, G. L. Klimchitskaya and Yu. V. Pavlov, for the invitation to participate and submit their works to this high-level International meeting.

\end{document}